\documentclass[preprint]{emulateapj}

\usepackage{threeparttable}
\usepackage[backref,breaklinks,colorlinks,citecolor=blue]{hyperref}

\shorttitle{Torque Variability in 1E$\;$1048.1$-$5937}
\shortauthors{Archibald et al.}

\begin{document}

\title{Repeated, Delayed Torque Variations Following X-ray Flux Enhancements in the Magnetar 1E$\;$1048.1$-$5937}

\author{R. F. Archibald\footnotemark[1], V. M. Kaspi\footnotemark[1], C.-Y. Ng\footnotemark[2], P. Scholz\footnotemark[1],   A. P. Beardmore\footnotemark[3], N. Gehrels\footnotemark[4], \& J. A. Kennea\footnotemark[5]}
\begin{abstract}
We report on two years of flux and spin evolution monitoring of 1E 1048.1$-$5937, a 6.5-s X-ray pulsar identified as a magnetar.  Using {\it Swift} XRT data, we observed an X-ray outburst consisting of an increase in the persistent 1--10 keV flux by a factor of 6.3$\pm$0.2, beginning on 2011 December 31 (MJD 55926).
Following a delay of $\sim100$ days, the magnetar entered a period of large torque variability, with $\dot{\nu}$ reaching a factor of $4.55\pm0.05$ times the nominal value, before decaying in an oscillatory manner over a time scale of months. We show by comparing to previous outbursts from the source that this pattern of behavior may repeat itself with a quasi-period of $\sim1800$ days. We compare this phenomenology to periodic torque variations in radio pulsars, finding some similarities which suggest a magnetospheric origin for the behavior of 1E 1048.1$-$5937. 

\end{abstract}
\section{Introduction}

 \footnotetext[1]{Department of Physics, McGill University, Montreal QC, H3A 2T8, Canada}
 \footnotetext[2]{Department of Physics, The University of Hong Kong, Pokfulam Road, Hong Kong}
 \footnotetext[3]{Department of Physics and Astronomy, University of Leicester, University Road, Leicester LE1 7RH, UK}
 \footnotetext[4]{Astrophysics Science Division, NASA Goddard Space Flight Center, Greenbelt, MD 20771, USA}
 \footnotetext[5]{Department of Astronomy and Astrophysics, 525 Davey Lab, Pennsylvania State University, University Park, PA 16802, USA}

The X-ray pulsar 1E$\;$1048.1$-$5937 is part of a small class of neutron stars known as magnetars.
Magnetars are characterized by their long ($\sim2-12\;$s) spin periods, extremely high spin-inferred magnetic fields ($\sim 10^{14}$ G), and X-ray luminosities which can greatly exceed the energy available from spin-down alone.
In addition, magnetars have periods of intense activity in which their X-ray flux can increase by orders of magnitudes and decay on a timescale of months. During such outbursts they can emit short ($\sim100\;$ms) hard X-ray bursts.  
The magnetar model \citep{1992ApJ...392L...9D, 1995MNRAS.275..255T, 1996ApJ...473..322T} was developed to explain the behavior of two classes of sources -- the Anomalous X-ray Pulsars (AXPs) and the Soft Gamma-ray Repeaters (SGRs). 
In this model, magnetars are powered by the decay of their magnetic fields which heats the stellar interior and causes internal stresses on the stellar crust which occasionally yields.
1E$\;$1048.1$-$5937 was the first AXP to show SGR-like bursts, helping to unify these two classes, as uniquely predicted by the magnetar model \citep{2002Natur.419..142G}.

1E$\;$1048.1$-$5937 was monitored regularly with the {\it Rossi X-ray Timing Explorer} ({\it RXTE}) from 1998 until its decommissioning in December of 2011 \citep{DibKaspi2013}.
During this monitoring, the source exhibited three long-term flux outbursts; one in 2001, followed by a second in 2002, and a third in 2007 \citep{2009ApJ...702..614D,DibKaspi2013}.
The first flux outburst was accompanied by SGR-like bursts from the source \citep{2002Natur.419..142G}.
Following both the second and third outbursts, order-of-magnitude variations in $\dot{\nu}$ were reported, but their origin was a mystery \citep{2004ApJ...609L..67G, 2009ApJ...702..614D,DibKaspi2013}.

Here we report an additional flux outburst in 1E$\;$1048.1$-$5937 as observed in X-ray timing observations obtained using the {\it Swift} X-ray Telescope in December of 2011.  
We show that again, roughly 100 days following the outburst, the pulsar's spin-down rate began showing large variations that are still on-going. 
We also show evidence of a quasi-periodicity in the torque during these increased spin-down periods.
This strongly suggests that such outbursts and long-term torque changes are causally related, and repeatable in this source. 
In addition, we report on a radio non-detection of the source during this torque enhanced period.

\section {Observations and Analysis}
\subsection{Swift XRT}
In July 2011, we began a monitoring campaign with the {\it Swift}  X-Ray Telescope (XRT) \citep{2005BurrowsSWIFT} of 1E$\;$1048.1$-$5937, along with five other magnetars.
This campaign is a continuation of a long-term monitoring of magnetars conducted with {\it RXTE}  \citep{DibKaspi2013}.
The {\it Swift} XRT is a Wolter-I telescope with an {\it XMM-Newton} EPIC-MOS CCD22 detector, sensitive in the $0.5-10\;$keV range.
The XRT was operated in Windowed-Timing (WT) mode for all observations.
This gave a time resolution of $1.76\;$ms. 
Observations, typically $1.5\;$ks long, were taken in groups of three, with the first two observations within 8 hours of each other and the third a day later.
This observation strategy was adopted due to the source's prior unstable timing behavior, where maintaining phase coherence using a longer cadence was only possible for several-month intervals \citep{2001ApJ...558..253K, 2009ApJ...702..614D}.
In all, 188 observations totalling $\sim$300 ks of observation time were analyzed.  
These observations are summarized in Table~\ref{ObsTable}.

\begin{table*}
\small
\centering
\centering
  \begin{threeparttable}
\caption{Summary of observations of 1E$\;$1048.1$-$5937 used in this work}
\label{ObsTable}
\begin{tabular}{ c c c c c c c}
\hline
\hline
Telescope & Target ID & Observation Dates & Typical Separation & Exposure & No. Obs.\\
  &   &   & (days)& (ks) &  \\
\hline
{\it Swift} XRT & 31220	&	2011-07-26 to 2013-10-16 & 14\tnote{1}	&	1.5\tnote{2}	& 188\\
{\it Chandra} ACIS-S & 14139	&	2012-02-23  & N/A	&	6	& 1\\
{\it Chandra} ACIS-S & 14140	&	2012-04-10  & N/A	&	12	& 1\\
ATCA & N/A	&	2013-03-08  & N/A	&	18 & 1\\
\hline
    \end{tabular}
    \begin{tablenotes}
  \item[1]{This separation was shortened to every 7 days near the maximal torque variations. After each separation, 3 closely spaced observations were taken. See Section$\;$\ref{sec:timing} for details.}
  \item[2]{This is the typical exposure time. Individual observations ranged from 1.1 ks to 7 ks.}
    \end{tablenotes}
  \end{threeparttable}
\end{table*}

Level 1 data products were obtained from the HEASARC \emph{Swift} archive, reduced using the $xrtpipeline$ standard reduction script, and barycentered to the location  of 1E$\;$1048.1$-$5937, $RA= 10^h50^m07.13^s$, $DEC=-59^\circ  53' 23.3''$ \citep{2002ApJ...579L..33W} using HEASOFT $v6.16$.
Individual exposure maps, spectrum, and ancillary response files were created for each orbit and then summed.
We selected only Grade 0 events for spectral fitting as higher Grade events are more likely to be caused by background events \citep{2005BurrowsSWIFT}.

To investigate the flux and spectra of 1E$\;$1048.1$-$5937, a 20-pixel radius circular region centered on the source was extracted.
As well, an annulus of inner radius 40-pixel and outer radius of 60-pixels centered on the source was used to extract background events.   

\subsection{Timing Analysis}
\label{sec:timing}
Barycentered events were used to derive a pulse time-of-arrival (TOA) for each observation. 
The TOAs were extracted using a Maximum Likelihood (ML) method as described in  \cite{2009LivingstoneTiming} and \cite{2012ApJ...761...66S}.
The ML method compares a continuous model of the pulse profile to the profile obtained by folding a single observation.
The template was derived from taking aligned profiles of all the pre-outburst {\it Swift} XRT observations and creating a profile composed of the first five Fourier components.

These TOAs were fitted to a timing model in which the phase as a function of time $t$ can be described by a Taylor expansion:
\begin{equation}
\phi(t) = \phi_0+\nu_0(t-t_0)+\frac{1}{2}\dot{\nu_0}(t-t_0)^2+\frac{1}{6}\ddot{\nu_0}(t-t_0)^3+\cdots
\end{equation}
where $\nu$ is the rotational frequency of the pulsar.
This was done using the TEMPO2 \citep{2006MNRAS.369..655H} pulsar timing software package. 

\begin{figure*}
\centering
\includegraphics[width=\textwidth]{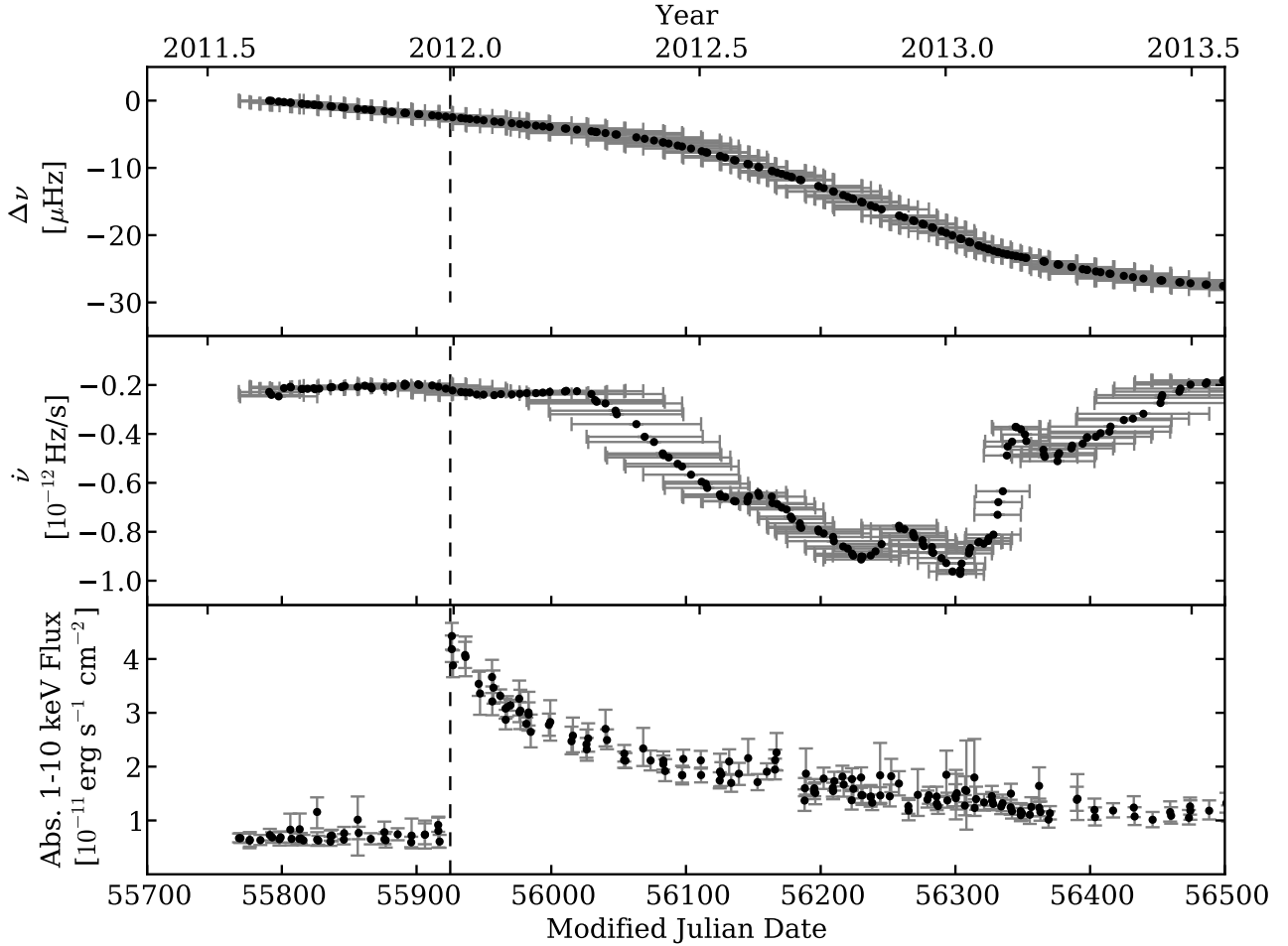}
\caption[Frequency panels]{Short-term timing parameter evolution of 1E$\;$1048.1$-$5937 surrounding the December, 2011 outburst. The top panel shows $\Delta\nu$ from the start of the {\it Swift} monitoring. The second panel shows $\dot{\nu}$.
In the top two panels, the horizontal error bars indicate the epoch over which $\nu$ and $\dot{\nu}$ were fit. 
 The bottom panel shows the total absorbed 1--10$\;$keV flux. The vertical dashed line indicates the start of the flux outburst.}
\label{freqpanels}
\end{figure*}

To ensure that phase-coherence was maintained over the periods of extreme torque variation, overlapping short-term ephemerides spanning 50 to 100 days were created, and TEMPO2 was used to extract pulse numbers.
Overlapping segments were compared to ensure that the same number of phase turns existed in overlapping segments between any two consecutive observations.
Each of these short, overlapping segments was fitted using TEMPO2 to a timing solution with just $\nu$ and $\dot{\nu}$, the results of which are presented in Figure~\ref{freqpanels}.

\begin{figure}
\centering
\includegraphics[width=\columnwidth]{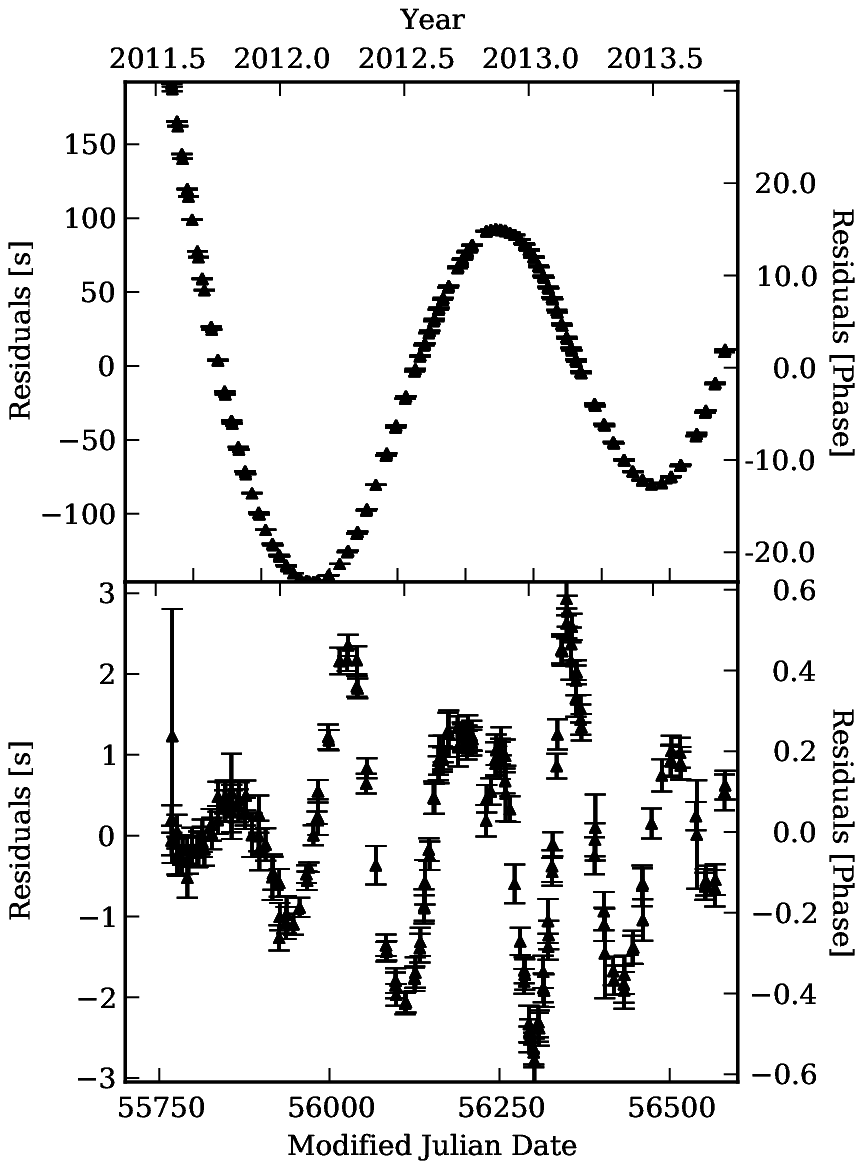}
\caption[Timing Residuals]{Timing residuals of 1E$\;$1048.1$-$5937 from MJD 55768-56581. The top panel shows residuals with only $\nu$ and $\dot{\nu}$ fitted. The bottom panel shows the residuals with 9 frequency derivatives fitted.}
\label{fig:resplot}
\end{figure}

This also allowed the extraction of an absolute pulse number for each TOA, allowing the fitting of one phase-coherent solution for the entire data set.
This solution is presented in Table~\ref{tab:res},  with the residuals in Figure~\ref{fig:resplot}.
Note that this solution is not a complete description of the spin of the source, as can be seen by the substantial residuals in Figure~\ref{fig:resplot}, and by the high $\chi^2/dof$ in Table~\ref{tab:res}.

\begin{table}
\centering
\centering
  \begin{threeparttable}
\caption{Timing Parameters for 1E$\;$1048.1$-$5937}
\label{tab:res}
\begin{tabular}{ c c }
\hline
\hline  
\rule{0pt}{3ex}
RAJ                      & 10:50:07.13\\          
\rule{0pt}{3ex}
DECJ                     & $-$59:53:23.3\\    
\rule{0pt}{3ex}
MJD Range               & 55768-56581  \\  
\rule{0pt}{3ex}
Epoch (MJD)             & 56000\\   
\rule{0pt}{3ex}  
$\nu$ (s$^{-1}$)        & $0.154 782 124(9)$\\  
\rule{0pt}{3ex}              
$\frac{d\nu}{dt}$(s$^{-2}$)               & $-2.43(2)\times10^{-13}$\\              
\rule{0pt}{3ex}  
$\frac{d^2\nu}{dt^2}$(s$^{-3}$)            & $-1.62(8)\times10^{-20}$\\             
\rule{0pt}{3ex}  
$\frac{d^3\nu}{dt^3}$ (s$^{-4}$)           & $-3.6(2)\times10^{-27}$\\            
\rule{0pt}{3ex}  
$\frac{d^4\nu}{dt^4}$ (s$^{-5}$)           & $-1.0(6)\times10^{-34}$\\           
\rule{0pt}{3ex}  
$\frac{d^5\nu}{dt^5}$ (s$^{-6}$)           & $9(2)\times10^{-41}$\\          
\rule{0pt}{3ex}  
$\frac{d^6\nu}{dt^6}$ (s$^{-7}$)           & $5(2)\times10^{-48}$\\         
\rule{0pt}{3ex}  
$\frac{d^7\nu}{dt^7}$ (s$^{-8}$)           & $-2(1)\times10^{-54}$\\        
\rule{0pt}{3ex}  
$\frac{d^8\nu}{dt^8}$ (s$^{-9}$)           & $-1(2)\times10^{-61}$\\   
\rule{0pt}{3ex}  
$\frac{d^9\nu}{dt^9}$ (s$^{-10}$)          & $3(2)\times10^{-68}$\\          
\rule{0pt}{3ex}
RMS Residual (s)        & 1.28\\                             
\rule{0pt}{3ex}
$\chi^2$/dof            & 10750.85/177\\

\hline
    \end{tabular}
    \begin{tablenotes}
    \item{ All errors are TEMPO2 reported 1$\sigma$ errors.}    
    \end{tablenotes}
  \end{threeparttable}
\end{table}

As is evident in Figure~\ref{freqpanels}, at the time of the flux outburst, we find no evidence for a glitch in $\nu$.
The data, however, are consistent with a change in the spin-down rate, $\Delta\dot{\nu}\sim 4\times10^{-14}\;$Hz s$^{-1}$, a $\sim10\%$ change in torque.
The exact amplitude of this change depends strongly on the length of the data span fit, and we are therefore unable to constrain this $\Delta\dot{\nu}$ to a single sudden event.    

Approximately 100 days following the peak of the flux outburst, the magnitude of $\dot{\nu}$ began to increase at a rate of $\sim-3\times 10^{-14}\;$Hz s$^{-1}$ day$^{-1}$ for the following 200 days, as seen in the middle panel of Figure~\ref{freqpanels}.
The spin-down rate continued to fluctuate on weeks to months timescales, with $\dot{\nu}$ changing by up to a factor of 5 in this time frame.

\subsection{Flux and Pulse Profile Evolution}
\label{sec:flux}
Spectra were extracted from the selected regions using {\tt extractor}, and fit using {\tt XSPEC} package version 12.8.2\footnote{http://xspec.gfsc.nasa.gov}.
The spectra were fit with a photoelectrically absorbed power law. 
We chose to use a single power law in place of the more commonly used power law and blackbody as for a given short observation, the statistics did not warrant a two-component model.  
Photoelectric absorption was modeled using {\tt XSPEC} {\tt tbabs} with abundances from  \cite{2000ApJ...542..914W}, and photoelectric cross-sections from  \cite{1996ApJ...465..487V}.
A single $N_H$ was fit to all pre-outburst spectra which yielded a best-fit value of $N_H=(1.98\pm 0.08)\times 10^{22}\;$cm$^{-2}$. 
For all the fluxes shown in Figures~\ref{freqpanels}, $N_H$ was held constant at this value while fitting the spectra. 
Co-fitting all pre-burst observations (MJDs 55768-55917) yielded $\Gamma=3.04\pm0.07$ with a 1--10$\;$keV absorbed flux of $7.0^{+0.1}_{-0.2}\times10^{-12}$erg\,cm$^{-2}$s$^{-1}$, with $\chi^2/dof= 505.11/501$.

\begin{figure*}
\centering
\includegraphics[width=\textwidth]{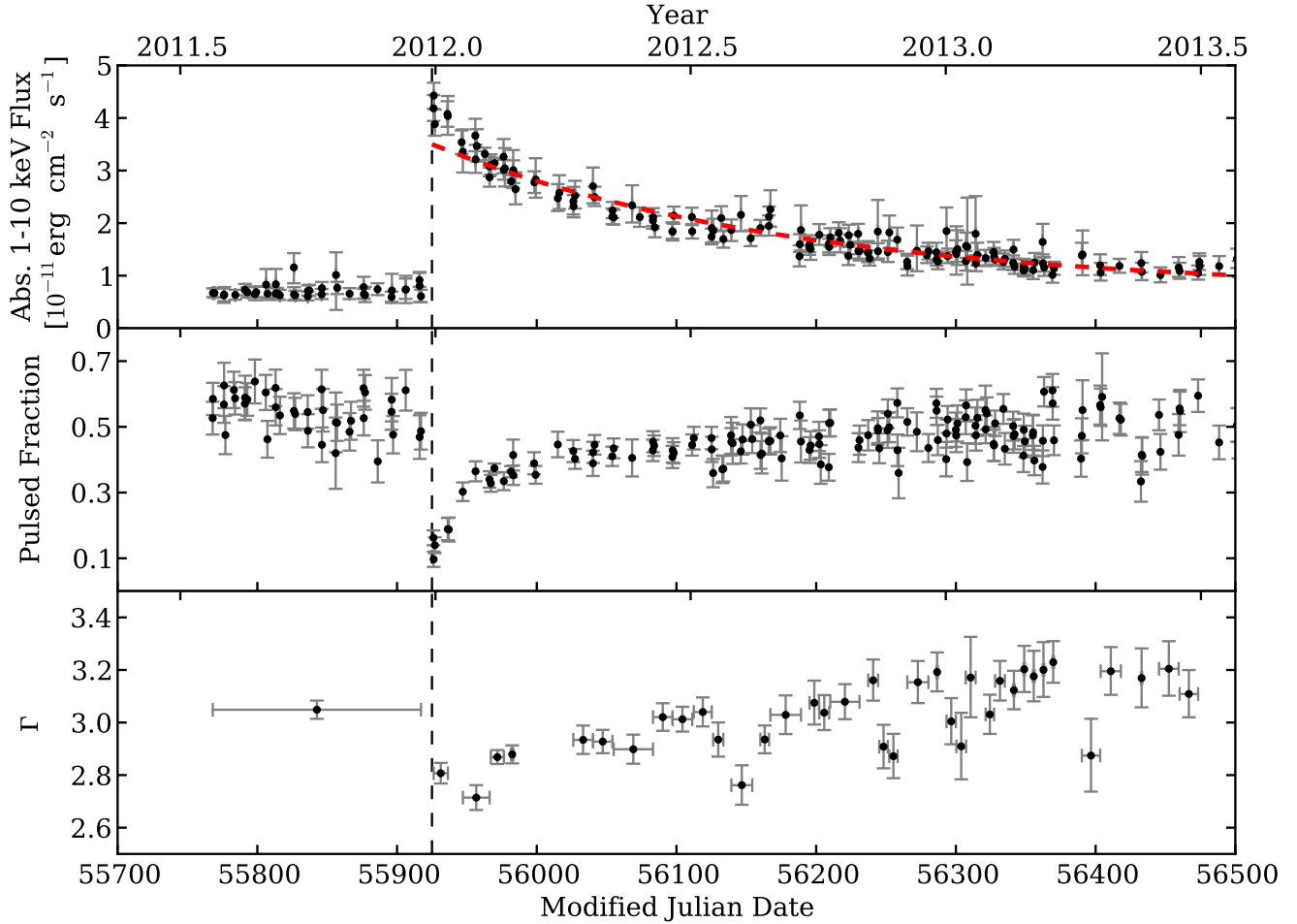}
\caption[Flux Plot]{Flux evolution of 1E$\;$1048.1$-$5937. The top panel shows the total absorbed 1--10 keV  flux, with the dashed line the fit to the post-burst flux decay, $F=[0.7^{+0.1}_{-0.2}+(2.8\pm0.3)e^{-(t-t_0)/(260\pm30)}] \times10^{-11}\;$erg s$^{-1}$cm$^{-2}$. The middle panel shows the RMS pulsed fraction, and the bottom panel, the power law index $\Gamma$, fit to the epochs indicated by the horizontal error bars. The vertical dashed line indicates the start of the flux outburst.}
\label{fluxpanels}
\end{figure*}

On MJD 55926 (2011 December 31) the measured 1--10 keV total absorbed flux increased sharply to ($4.4\pm0.15)\times10^{-11}$ erg cm$^{-2}$s$^{-1}$, a factor of 6.3$\pm$0.2 increase, as seen in Figure~\ref{freqpanels}.
The source also became harder, with $\Gamma=2.75\pm0.06$ for this observation.
Between the set of observations on MJD 55926 (2011 December 31) and the next set on MJD 55936 (2012 January 6), the flux fell to ($3.8\pm0.2)\times10^{-11}$ erg cm$^{-2}$s$^{-1}$.
After this initial decay, the 1--10$\;$keV flux decay is well described ($\chi^2/dof=132.3/132$) by an exponential decay: $F=[0.7^{+0.1}_{-0.2}+(2.8\pm0.3)e^{-(t-t_0)/(260\pm30)}] \times10^{-11}\;$erg s$^{-1}$cm$^{-2}$ where $t$ and $t_0$ are in units of days. $t_0$ was held fixed at MJD 55926, the peak of the outburst.
The pulsed fraction displayed in Figure~\ref{fluxpanels} is the root mean squared (RMS) pulsed fraction, as described in \cite{2004ApJ...605..378W}.
The clear correlation between the power law index and the measured 1--10 keV total absorbed flux apparent in Figure~\ref{fluxpanels} is typical for magnetar outbursts, see \cite{2011ApJ...739...94S}.  

For the 9 days between the prior {\it Swift} monitoring observation on MJD 55917 and the observation on MJD 55926 which had an enhanced flux, we detect no signficant emission in the {\it Swift} Burst Alert Telescope from the direction of the source.  We can place an upper limit on the 15-50$\;$keV emission of $7.5\times 10^{-5}$ counts$\;$s$^{-1}$\,cm$^{-2}$ ($\sim7\times10^{-12}$ erg cm$^{-2}$s$^{-1}$).\footnote{Hans Krimm, private communication.}
This limit indicates that the majority of the energy of this outburst is in the long exponentially decaying tail, rather than in a missed sharp burst.

In Figure~\ref{fig:PF_nudot}, the pulsed component of the flux of 1E$\;$1048.1$-$5937 is presented for the three long-term flux outbursts observed from this source.
The pulsed flux does not follow the fast rise seen in the total flux; instead we see a several weeks long rise in the pulsed flux before it begins to decay.
The {\it Swift} observations, where we can measure both pulsed and total flux, suggest that the slow rise times in the prior outbursts observed with {\it RXTE} \citep{2009ApJ...702..614D} are a result of {\it RXTE} being only sensitive to the pulsed flux from the source.
An anti-correlation between total flux and pulsed fraction has been reported for this source during previous outbursts \citep{2005A&A...437..997T, 2008ApJ...677..503T}. 
In particular \cite{2008ApJ...677..503T} fit the relation to a power law $P_F\propto F_x^n$ with index $-0.46\pm0.02$.
While this provides an adequate description of the pulsed fraction at most epochs, it overestimates the pulse fraction at the peak epoch by a factor of $\sim3$.

\begin{figure*}
\centering
\includegraphics[width=120mm]{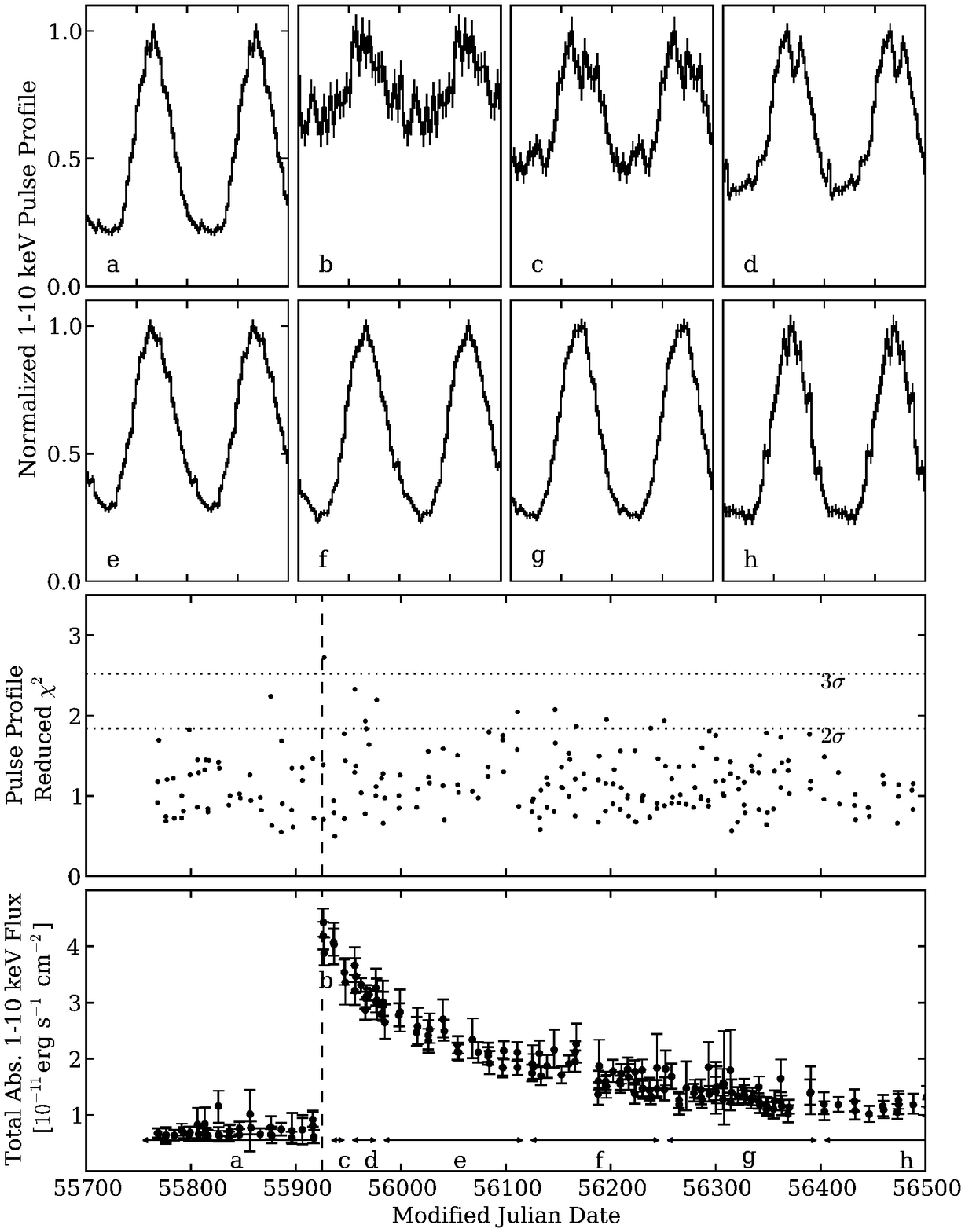}
\caption[Pulse profiles]{Normalized 1-10 keV pulse profiles for 1E$\;$1048.1$-$5937 around the 2012 outburst. 
The lettered panels at the top are the normalized background-subtracted pulse profiles summed over the time spans indicated by the arrows in the bottom panel. The profile b is from the observations at peak flux on MJD 55926.
The middle panel shows the reduced $\chi^2$ statistic calculated after subtracting the scaled and aligned profiles of the individual observations from a high signal-to-noise template. The lower and upper dotted lines correspond to $2\,\sigma$ and $3\,\sigma$ significance, respectively.
In the lower two panels, the vertical dashed lines indicated the start of the flux outburst.
To calculate the flux, $N_H$ was held constant at $1.98 \times 10^{22}\;$cm$^{-2}$; see \S~\ref{sec:flux}.}
\label{profs}
\end{figure*}

To monitor for changes in the pulse profile of 1E$\;$1048.1$-$5937, we created a high signal-to-noise 32-bin template by aligning all quiescent {\it Swift} XRT observations using the TOA offsets from the ML procedure described above.
For each observation, a phase-aligned profile was created using the current timing ephemeris.
The best-fit DC level was then subtracted from each profile, and the latter was scaled to match the template using a multiplicative  scaling factor which minimized the reduced $\chi^2$ of the difference between the scaled profile and the template.
The reduced $\chi^2$ values are presented in Figure~\ref{profs}, as well as the normalized 1-10$\;$keV pulse profiles surrounding the outburst. 
We note that the first observation following the outburst is inconsistent with the template profile at the $3\,\sigma$ level.
All other individual profiles are consistent with the template at the $3\,\sigma$ level, however there is a marked increase in the average reduced $\chi^2$ at the time of the flux increase.

Motivated by the increase in the average reduced $\chi^2$ near the outburst, we looked for lower-level changes in the pulse profiles. Nearby observations were combined to create higher signal-to-noise profiles which can be seen in Figure~\ref{profs}.
With these higher signal-to-noise profiles, we note low-level changes in the pulsed profiles, however the dominant change apparent is the large change in the pulsed fraction.

\subsection{{\it Chandra}}
Following the detection of the flux increase with the {\it Swift} monitoring campaign, a set of two {\it Chandra} Target of Opportunity observations were triggered in Continuous Clocking mode using {\it ACIS-S}. The data were processed using CIAO $4.5$ and CALDB $4.5.7$. Data were reprocessed using $chandra\_repro$ and barycentered using the $axbary$ tool.

To study the flux and spectra, a 6\arcsec$\;$ long-strip centered on the source was extracted as well as a background strip of total length 400\arcsec$\;$ located away from the source.   
For the {\it Chandra} data, the two spectra were co-fit with a single $N_H$, with all other parameters free.
The results of this fit can be seen in Table~\ref{ChanTable}.
We note that the $N_H$ presented here appears inconsistent with that of \cite{2004ApJ...608..427M} and \cite{2008ApJ...677..503T}.
This is due to a difference in the model used for photometric absorptions, as well as the values used for solar abundances and cross sections.
We can obtain a $N_H$ consistent with the previously reported values by fitting with the solar abundances and cross sections used by Tam et al.

\begin{table*}
\centering
\centering
  \begin{threeparttable}
\caption{{\it Chandra} spectral fits of 1E$\;$1048.1$-$5937\tnote{1}.}
\label{ChanTable}
\begin{tabular}{ c c c c c c c}
\hline
\hline
Date & Exposure  & kT & $\Gamma$ & Abs. 1-10 keV Flux \\
  & (ks) & (keV) &  & ($10^{-12}$ erg cm$^{-2}$ s$^{-1}$)\\
\hline
2012-02-23 & $6$	&	$0.624\pm{0.007}$	& $2.47^{+0.17}_{-0.22}$ & $27.5^{+0.3}_{-1.7}$\\
2012-04-10 & $12$   	&	$0.573^{+0.06}_{-0.07}$	& $2.29^{+0.17}_{-0.22}$ & $23.4^{-0.4}_{-1.6}$\\
\hline
    \end{tabular}
    \begin{tablenotes}
  \item[1]{$N_H$ was co-fit to both observations simultaneously, yielding $N_H = 1.29^{+0.09}_{-0.1} \times 10^{22}$ cm$^{-2}$}
  \end{tablenotes}
  \end{threeparttable}
\end{table*}

The pulse profiles obtained from the {\it Chandra} data are consistent with those presented from the {\it Swift} data.

\subsection{Australia Telescope Compact Array}\label{obs:ATCA}
We performed a search for radio emission using the Australia Telescope Compact
Array (ATCA). The observation was made on 2013 Mar 4 in the 6A array
configuration with a total integration time of 5\,hr. The central frequency was
2.1\,GHz with a bandwidth of 2\,GHz. The flux density scale was set by
observations of the primary calibrator, PKS B1934$-$638. A secondary
calibrator, PKS B1049$-$53, was observed every 40\,min to determine the
antenna gains.

We carried out all data reduction with the MIRIAD
package\footnote{\url{http://www.atnf.csiro.au/computing/software/miriad/}}.
After standard flagging and calibration, an intensity map was formed
using the multi-frequency synthesis technique with natural weighting,
then deconvolved using a multi-frequency CLEAN algorithm (\texttt{mfclean}),
and restored with a Gaussian beam of FWHM $7.5\arcsec \times 5.1\arcsec$. The
final image has rms noise of 0.06\,mJy\,beam$^{-1}$.
We note that this is a few times higher than the theoretical limit of 
0.015\,mJy\,beam$^{-1}$, due to the sidelobe of an extended extragalactic
source G288.27$-$0.70  \citep{2007ApJ...663..258B} 10\arcmin\ to the
southwest. We found no source at the position of 1E$\;$1048.1$-$5937. This yields
a $3\sigma$ flux density limit of 0.2\,mJy at 2.1\,GHz.
This corresponds to a $L_{2.1GHz}/L_{1-10 keV}< 4\times10^{-7}$ at the epoch of the observation.
Magnetars are highly variable in the radio band; for example, in the case of the magnetar XTE$\;$J1810$-$197 $L_{1.4GHz}/L_{0.6-10 keV}$ varies from $7\times10^{-5}$ to less than $7\times10^{-7}$\citep{2007ApJ...663..497C, 2009A&A...498..195B}.
Other magnetars also show orders of magnitude variations in their radio luminosity, see \cite{2014ApJS..212....6O} and references therein.

Previous searches for radio emission from this source have also resulted in non-detections.
A $3\sigma$ flux density limit of 0.11\,mJy at 1.4\,GHz  \citep{2006MNRAS.372..410B} in 1999 was found when the source was spinning down steadily.
Also the source was not detected in a pulsation search $\sim$15 days after the 2007 glitch, for a period-averaged flux density limit  at 1.5 GHz of 0.1\,mJy \citep{2007ATel.1056....1C}.

\section{Discussion}
\label{sec:discussion}

We have presented {\it Swift} XRT and {\it Chandra ACIS-S} observations of 1E$\;$1048.1$-$5937 which show a sudden factor of $6.3\pm0.2$ increase in the total $1-10\;$keV X-ray flux on MJD 55926 (2011 December 31).
This flux showed different evolution in the pulsed versus total flux, with the total flux rising sharply, and the pulsed flux showing a slow rise on a $\sim$weeks timescale.
This flux increase was followed, after a $\sim100$ day delay, by a months-long torque increase.
We also report an upper limit of 0.2$\;$mJy at 2.1 GHz using the Austrailia Telescope Compact Array during this  period of torque enhancement.

1E 1048.1$-$5937 displays a stong anti-correlation between the total X-ray flux, and the pulsed fraction, as can be seen in Figure~\ref{fluxpanels}.
Similar anti-correlations between X-ray flux and pulsed fraction have been reported during other several magnetar outbursts \citep[e.g.][]{2007ApJ...664..448I, 2011ApJ...729..131N,2013ApJ...770...65R}.
This suggests that at the time of outbursts, energy is being injected into the magnetar isotropically, eg. across the entire surface or magnetosphere rather than being injected at one point.

\begin{figure*}
\centering
\includegraphics[width=\textwidth]{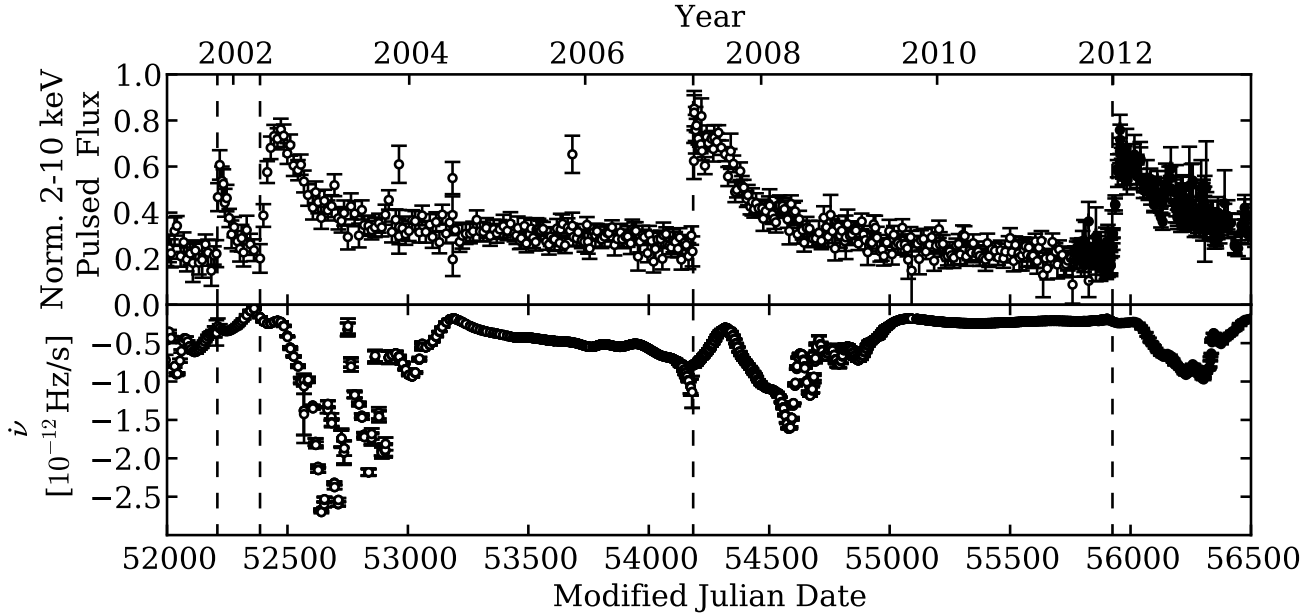}
\caption[Pulsed Flux Nudot]{Pulsed flux and $\dot{\nu}$ evolution for 1E$\;$1048.1$-$5937. The top panel shows the normalized 2--10 keV pulsed flux. The bottom panel shows $\dot{\nu}$. The dashed lines indicate the start times of the pulsed flux outbursts. In both panels,  {\it RXTE} data are hollow points, and {\it Swift} data are solid points. {\it RXTE} data are from \citet{DibKaspi2013}.}
\label{fig:PF_nudot}
\end{figure*}

1E 1048.1$-$5937 has shown similar behavior in the past, with large pulsed flux outbursts in 2001, 2002, and 2007 \citep{2004ApJ...609L..67G, 2009ApJ...702..614D,DibKaspi2013}. All of these pulsed flux outbursts have similar increases in the 2-10 keV pulsed flux with a factor of $2.3\pm0.2$, $2.9\pm0.1$, $3.1\pm0.1$, and $3.3\pm0.2$ for the four outbursts.
 We note that for the three outbursts which are followed by extreme torque variation, the pulsed flux increases are consistent with each other, and are larger than in the outburst in 2001.
In Figure~\ref{fig:PF_nudot}, we show these outbursts along with the most recent flux outburst reported here.
For the 4500 days of data examined in this work, $42\pm6\;\%$ of the pulsed X-ray emission from the source has been due to these pulsed flux flares, and the remainder due to its baseline flux.

The possible hint of periodicity in the three largest flux enhancements is not strict: the times between the pulsed flux outbursts are $180\pm10$, $1800\pm10$, and $1740\pm10$ days, respectively. This variation by $\sim$3\% after just 3 cycles immediately precludes a binary-companion-related origin, wherein a strict periodicity would be expected. 
On the other hand, the reality of the possible periodicity is supported by its being observed independently in the flux evolution and in the torque evolution.

\begin{figure*}
\centering
\includegraphics[width=\textwidth]{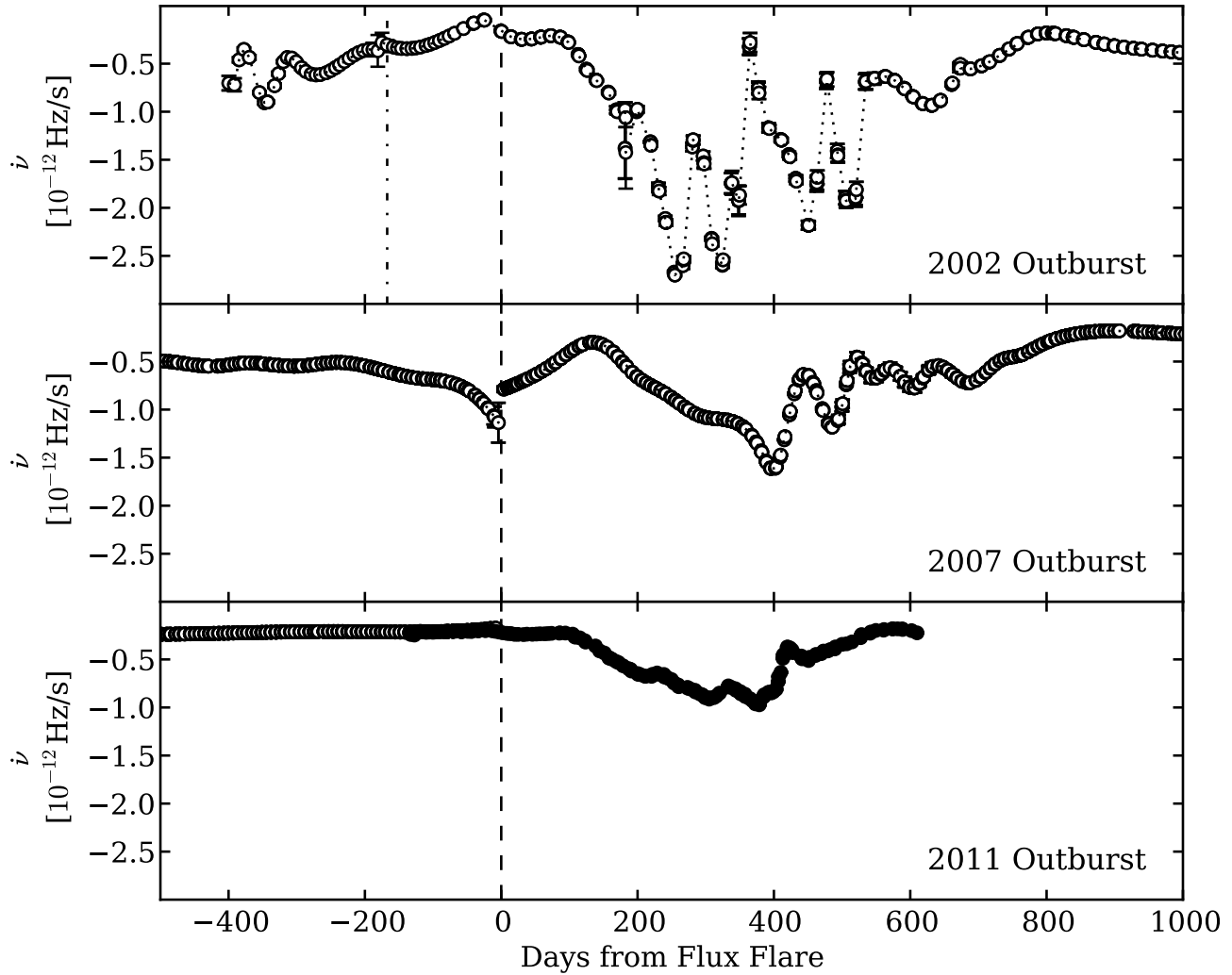}
\caption[Nudot Zoom]{$\dot{\nu}$ evolution for 1E$\;$1048.1$-$5937 following each flux increase. The dashed lines indicate the start times of the pulsed flux outbursts, with the dot-dash line in the top panel indicating the precursor flare in 2001. {\it RXTE} data are hollow points, and {\it Swift} data are solid points. {\it RXTE} data are from \citet{DibKaspi2013}.}
\label{fig:nudot}
\end{figure*}

\begin{figure}
\centering
\includegraphics[width=\columnwidth]{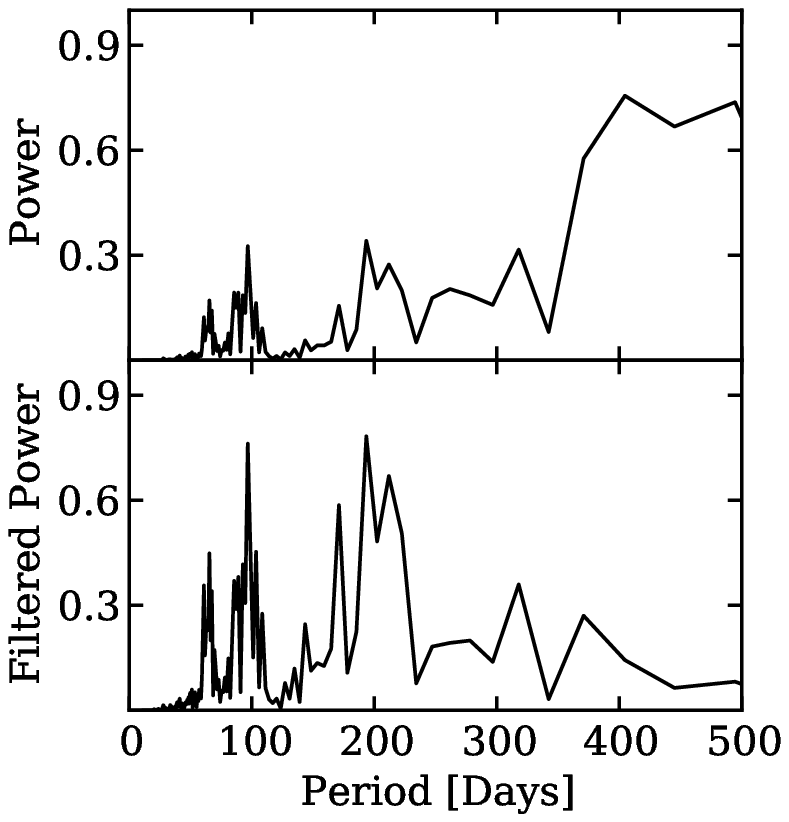}
\caption[Torque Power]{Normalized power spectrum of $\dot{\nu}$ for 1E$\;$1048.1$-$5937 over 12 years. The top panel is red-noise unfiltered, and the bottom panel has been passed through a high-pass filter. See Section~\ref{sec:discussion} for details. }
\label{Torque_Power}
\end{figure}

The extreme variation in $\dot{\nu}$ has become smaller with each subsequent outburst, reaching minimum values of $(-2.70\pm0.01)\times 10^{-12}$ Hz s$^{-1}$, $(-1.61\pm0.01)\times 10^{-12}$ Hz s$^{-1}$, and $(-0.97\pm0.01)\times 10^{-12}$ Hz s$^{-1}$ for the three outbursts, respectively. 
For comparison, the longest pseudo-stable spin-down rate is $-2.2\times 10^{-13}$ Hz s$^{-1}$.
In all three cases $\dot{\nu}$ does not evolve symmetrically, rather it seems to increase nearly monotonically then decay back to the nominal spin-down value in an oscillatory manner.
This can seen more clearly in Figure~\ref{fig:nudot}, where the $\dot{\nu}$ variations are shown in days from the detected flux increase. 
To characterize the oscillations visible in $\dot{\nu}$ following the pulsed flux increases, we calculated a power spectrum of the observed $\dot{\nu}$, shown in Figure~\ref{Torque_Power}.  We note broad peaks at $64\pm4$, $96\pm10$, and $200\pm 20$ days.

The origin of the observed possibly quasi-periodic flux enhancements and subsequent spin-down rate variations is puzzling.
A periodicity in spin period on comparable time scales
was predicted by \citet{1999ApJ...519L..77M} for this source in particular; 
this was suggested to result from the deformation of the neutron star from sphericity
due to the high $B$ field, causing Eulerian precession.  
\citet{2000ThompsonSpinDown} also considered magnetar precession; they and \citet{1999ApJ...519L..77M}
predicted strictly periodic behavior, not the more intermittent nature
observed in 1E 1048.1$-$5937 wherein the pulsar goes for several years
with relatively stable spin down.  Apart from
the only approximately periodic nature of the behavior not agreeing with the 
precession models, X-ray outbursts are similarly not predicted in such models, and certainly no
clear phase delay in a quasi-cycle between spin and flux behavior was
predicted. 
Moreover, \citet{1977ApJ...214..251S} suggested that neutron-star precession with period
$\tau_p$ requires that the moment of inertia of any pinned crustal superfluid,
such as that invoked to explain glitches like those seen ubiquitously
in magnetars \citep[e.g.][]{2008Dib,DibKaspi2013}, not exceed a fraction
$P/\tau_p$ of that of the neutron star -- far smaller than has been
inferred for both magnetars and radio pulsars.
No such long-term quasi-periodicities have been predicted or discussed in any
other magnetar paper, nor in any disk model, to our knowledge.

However we note that quasi-periodicities in spin-down rate have been reported
on very similar time scales in radio pulsars by \citet{2010Sci...329..408L} (and see also \cite{2006Sci...312..549K}).  Those authors
reported on long-term monitoring of 366 radio pulsars using the Lovell Telescope
at Jodrell Bank, finding at least 18 sources to show clear, substantial and generally
quasi-periodic changes in $\dot{\nu}$ in data sets over 20 years long.  The quasi-periodicities
ranged in time scale from months to years depending on the source, and in at least
six cases, were clearly correlated with changes to the radio pulsars' average profiles.
Notably, \citet{2010Sci...329..408L} observed that in those six cases, the $\dot{\nu}$ and the
pulse morphology, appeared in two or more distinctly preferred `states.'  Because
of the strong correlation between the spin-down evolution and radio pulsar shape
changes, \citet{2010Sci...329..408L} concluded the behavior must have its origin in the stellar
magnetosphere, and observationally ruled out previous suggestions that it could be free precession
\citep{2000Natur.406..484S}, which again, appeared challenging to understand theoretically \citep{1977ApJ...214..251S}.
Moreover, apparent evolution in the magnetic field structure has also been seen in the Crab pulsar
\citep{2013Sci...342..598L}, further suggesting this behavior is ubiquitous in radio pulsars.

In 1E 1048.1$-$5937, we find behavior that appears in many way similar to that of the radio
pulsars studied by \cite{2010Sci...329..408L}:  quasi-periodic spin-down on time scales of several
years, with two apparently distinct `states,' one fairly steady, and one quasi-oscillatory.
These are clearly correlated with radiative changes, in this case in the form of enhanced
X-rays, whose origin can be explained by magnetospheric twists which result in enhanced
return currents that cause the observed spectrum to harden with increased flux
\citep{2002ApJ...574..332T,2007ApJ...657..967B,2009ApJ...703.1044B}.

The magnitude of the spin-down-rate changes in 1E 1048.1$-$5937 is much larger than
in the \citet{2010Sci...329..408L} objects. In those cases though typical values of
$\Delta \dot{\nu}/\dot{\nu}$ were $\sim$1\%, although a value of 45\% was observed in one source \citep[see also][]{2006Sci...312..549K}.
The latter is still much smaller than the factor of $> 10$ changes we have observed in 
1E 1048.1$-$5937 \citep{2004ApJ...607..959G}.  However given independent conclusions about the greatly enhanced activity
in magnetar magnetospheres, on the basis of their radiative behavior overall (bursts, long-term flux enhancements, etc), it is perhaps unsurprising that the magnitude of the analogous $\dot{\nu}$ effect in a 
magnetar should be much larger than that in conventional radio pulsars. 
This suggests there could be a greater tendency for higher-$B$ radio pulsars to show
magnetospheric evolution as manifested by radio pulse variations and/or spin-down fluctuations.
However, in the 18 pulsars discussed by \citet{2010Sci...329..408L}, there is no such correlation seen between magnetic field strength and $\dot{\nu}$ variation.

In the case of 1E 1048.1$-$5937, these epochs of extreme torque variability dominate the average spin-down of the magnetar. 
If we take the difference in the spin frequency at the start and end of the data presented in this work, MJD 52000-56500, we measure an average spin-down rate of approximately $-5.5\times10^{-13}\,$Hz$\,$s$^{-1}$.
This is a factor of $\sim2.5$ higher than the quiescent spin-down of $-2.2\times10^{-13}\,$Hz$\,$s$^{-1}$ measured during the least active period in 2010 and 2011. 

The apparent absence thus far of similar behavior in other magnetars could possibly be understood as being due to
the same mechanism that causes such a variety of periodicity time scales in the Lyne et al. sources.
Longer-term quasi-periodicities may eventually become apparent in other magnetars.
If radio pulsations were one day to be observed from 1E 1048.1$-$5937 in spite of
strong upper limits as reported both here and elsewhere \citep{2006MNRAS.372..410B,2007ATel.1056....1C}, we would predict
correlated changes with spin-down rate.  However, this may never be observed, if only due to
unfortunate beaming \citep[e.g.][]{2012ApJ...744...97L}.

The observed delay between the X-ray outbursts and $\dot{\nu}$ `state' changes
has already been considered by \citet{2007ApJ...657..967B}.  As also discussed by \citet{2010MNRAS.408L..41T} but in
the context of conventional radio pulsars, a time delay is expected between when the twist
forms near the neutron-star crust, and when it reaches the outer magnetosphere, where the
impact on the spin-down rate manifests.  This time delay is suggested by
\citet{2007ApJ...657..967B} to be due to the spreading time of twist current across magnetic field lines, a result of magnetospheric coronal resistivity, and its value is predicted to
be highly geometry-dependent.  This argues that the events that trigger
the flux enhancements in 1E 1048.1$-$5937 must consistently be in the same approximately 
region of the neutron-star surface, and likely far from the poles, where little delay is expected.
Likely then the monotonic decline in the amplitude of the torque variations seen in the
three cycles observed so far is a coincidence, and future cycles may yet again exhibit
order-of-magnitude torque changes.

Indeed if the apparent quasi-periodicity is real, we expect the next X-ray flux outburst cycle
to occur in late 2016, with the next cycle of $\dot{\nu}$ oscillations a few months later.
Future X-ray telescopes will easily observe this behavior, should it occur.
Moreover, continued long-term monitoring of other magnetars may yet reveal similar behavior,
though the Jodrell Bank work underscores the need for systematic monitoring over decades.

\section{Conclusions}
We have reported on long-term systematic monitoring of the magnetar 1E 1048.1$-$5937 using the {\it Swift} X-ray Telescope.  This monitoring has revealed:

\noindent
(1) Evidence for quasi-periodic X-ray outbursts, each of comparable
amplitude, roughly a factor of $\sim3$ above the typical pulsed level, with
approximate recurrence time scale $\sim$1800 days.  Considering
three events observed thus far, the difference in separation between the
first and second two was $\sim$3\% smaller than between the second
and third. 

\noindent
(2) Similar quasi-periodicity is seen in the evolution of $\dot{\nu}$ on the same
time scale; every $\sim$1800 days, $\dot{\nu}$ commences fluctuating
with quasi-oscillatory behavior on time scales of $\sim$100 days,
with amplitude changes as large as an order of magnitude, such that the
neutron star is always spinning down.  The fluctuations at each cycle
are not identical but are similar in time scale.

\noindent
(3) Though the flux and $\dot{\nu}$ both appear quasi-periodic with the
same repetition rate, there is a clear delay between the two such
that the $\dot{\nu}$ variations begin $\sim$100 days after
the flux outburst begins.

\noindent
(4)  The maximum amplitude of the $\dot{\nu}$ oscillations has declined monotonically
over the observed three cycles.

\noindent
(5)  The spin-down is relatively stable between episodes of $\dot{\nu}$
fluctuation.

\noindent
(6) Aside from the extreme variation in $\dot{\nu}$ always beginning after an X-ray outburst, there is no correlation betweem the X-ray luminosity and $\dot{\nu}$.

Although similar quasi-periodic radiative and torque behavior has not yet been reported in other magnetars, similar, though lower amplitude, behavior has been seen in radio pulsars \citep{2010Sci...329..408L,2006Sci...312..549K}, and appears to implicate processes in the stellar magnetosphere rather than, for example, precession.  Continued long-term monitoring of both source classes is warranted, to see if, for example, there is a greater tendency for such phenomena in more highly magnetized objects.

\section{Acknowledgements} We thank Jamie Stevens for carrying out the ATCA observations. The Australia Telescope Compact Array is part of the Australia Telescope National Facility which is funded by the Commonwealth of Australia for operation as a National Facility managed by CSIRO\footnote{\url{http://www.atnf.csiro.au/research/publications/Acknowledgements.html}}.
R.F.A. receives support from a Walter C. Sumner Memorial Fellowship.
V.M.K. receives support from an NSERC Discovery Grant and Accelerator Supplement, Centre de Recherche en Astrophysique du Quebec, an R. Howard Webster Foundation Fellowship from the Canadian Institute for Advanced Study, the Canada Research Chairs Program and the Lorne Trottier Chair in Astrophysics and Cosmology.
 We thank M. Lyutikov, D. Tsang, and K. Gourgouliatos for useful discussions. 
 We also thank an anonymous referee for comments which improved the manuscript.
 We acknowledge the use of public data from the {\it Swift} data archive. 
 This research has made use of data and software provided by the High Energy Astrophysics Science Archive Research Center (HEASARC), which is a service of the Astrophysics Science Division at NASA/GSFC and the High Energy Astrophysics Division of the Smithsonian Astrophysical Observatory.
 The scientific results reported in this article are based in part on observations made by the {\it Chandra} X-ray Observatory.
 This research has made use of CIAO software provided by the {\it Chandra} X-ray Center (CXC).
\bibliography{1048}{}
\bibliographystyle{apj}

\end{document}